\documentclass[12pt]{iopart}
\usepackage{epsf}
\begin{document}

\title[LC-oscillators using voltage-amplifiers]{LC sine-wave oscillators using general-purpose voltage operational-amplifiers}

\author{M. M. Jakas and F. Llopis}

\address{Departamento de F{\'i}sica Fundamental y Experimental, Electr\'onica y Sistemas, Universidad de La Laguna, 38205 Tenerife, SPAIN}

\ead{mmateo@ull.es and fllopis@ull.es}

\begin{abstract} It has been found that some text-books show LC-oscillators that may not work as assumed. Thus, the typical example showing a LC-oscillator driven by a voltage operational-amplifier is simply wrong. The difficulty stems from the fact that such oscillators are normally built to work with transconductance- not with voltage-amplifiers. Such a difficulty however, can be readily solved by connecting a resistor in series with the so-called frequency-determining network. 

\end{abstract}

\pacs{01.50.Pa}

\submitto{Physics Education}

\maketitle

\begin{figure}
\begin{center}
\epsfxsize=12cm\epsfbox{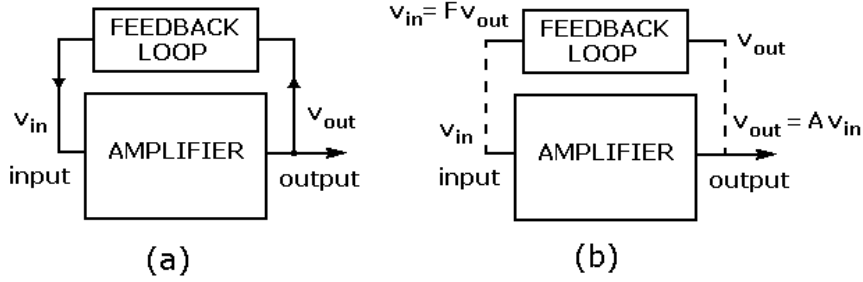}
\end{center}
\caption{\label{Fig1} The basic oscillator block diagram.  } 
\end{figure}

As is shown in text-books of elementary electronics, a sinusoidal oscillator circuit contains an amplifier stage and a feedback-loop as appears in Fig.\ref{Fig1} (a) \cite{Malik,Malvino}. This type of circuits are often analysed by using the voltage transfer-coefficient of the amplifier $A(\omega)$ and that of the feedback-loop $F(\omega)$, where $\omega$ is the frequency of the signal [see Fig.\ref{Fig1}(b)]. As one can readily see, the circuit is assumed to be capable of producing a self-sustained sine-wave signal if there is a frequency, $\omega_0$, for which $A(\omega_0)F(\omega_0)\geq$1. This is called the \emph{Barkhausen} condition after the German scientist Heinrich Georg Barkhausen (1881-1956). Since $A(\omega)$ and $F(\omega)$ may be complex numbers, the Barkhausen condition actually implies the following two equations
 
\begin{equation}
\label{Eq.1}
\Re{[A(\omega_0)F(\omega_0)]} \geq 1 \mbox{\hspace{1cm} and \hspace{1cm}}  \Im{[A(\omega_0)F(\omega_0)]} = 0 \, ,
\end{equation}

\noindent where $\Re{(X)}$ and $\Im{(X)}$ denote the real and imaginary part of $X$, respectively. 

For most amplifier stages, however, $A(\omega)$ has no imaginary part. As a matter of fact, $A(\omega)\approx A$ where $A$ may be assumed to be a real, negative or positive constant for an inverter or a non-inverter amplifier, respectively. Eqs.(\ref{Eq.1}), therefore, reduce to $\Im{[F(\omega_0)]}=0$ and  $A\Re{[F(\omega_0)]} \geq 1$. The equations above enable the calculation of the frequency of oscillation ($\omega_{0}$) and the minimum $|A|$ of the amplifier required to sustain the oscillation, i.e. $|A|_{min}=1/|\Re{[F(\omega_0)]}|$. 

It must be mentioned though, that $\Im{[F(\omega)]}$ can be forced to have a zero at a predetermined frequency, thanks to the fact that the feedback-loop contains a circuit known as the frequency-determining network (FDN). A FDN can be realized using a combination of resistors and capacitors as in the family of RC-oscillators or, inductors and capacitors as in the so-called LC-oscillators. Depending on the particular FDN the oscillator is often named after the person who designed it. 

Curiously enough, one can apply the analysis above to RC-oscillators, and both the oscillation frequency $\omega_{0}$ and $|A|_{min}$ are readily obtained. But when it comes to LC-oscillators some difficulties appear. This shortcoming is normally circumvented thanks to the fact that single-stages, either BJT or JFET amplifiers, are not \emph{voltage} but \emph{transconductance} amplifiers. Therefore, the analysis of such amplifiers has to be carried out in a different manner and, in the end, it all seems to work fairly well \cite{Malik}. However, in some text-books such a difficulty appear to be overlooked when a voltage operational-amplifier is proposed as an example of LC-oscillators \cite{Boylestad,REA}. This is definitely not correct and, although such an oscillator may work, in the sense that it may oscillate, it will not produce a sinusoidal signal and naturally, the oscillation frequency may not be necessarily the one predicted by the analysis above.

\begin{figure}
\begin{center}
\epsfxsize=5cm\epsfbox{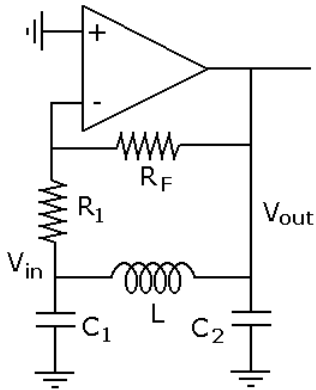}
\end{center}
\caption{\label{Fig3} The Colpitts oscillator as appears in some text-books. See for example Refs.\cite{Boylestad,REA}. } 
\end{figure}

In order to see this in more detail, one can study the LC-oscillator named after Colpitts that appears in Figure \ref{Fig3} \cite{Boylestad,REA}. For such a circuit the following results can be readily obtained,

\begin{equation}
\label{Eq.2}
A=-\frac{R_{F}}{R_1} \mbox{\hspace{1cm} and \hspace{1cm}} F(\omega) = \frac{R_1}{R_1(1-\omega^2C_{1}L)+j\omega L}\, ,
\end{equation}
   
\noindent where $j$ is the imaginary unit. 

As one can see, the Barkhausen criteria, i.e. $A(\omega)F(\omega)\geq $1, cannot be accomplished for any real, different-from-zero value of $\omega$. However, as was previously mentioned, there is nothing unexpected in these results since the FDN used in these circuits does not work with a voltage amplifier. 

\begin{figure}
\begin{center}
\epsfxsize=5cm\epsfbox{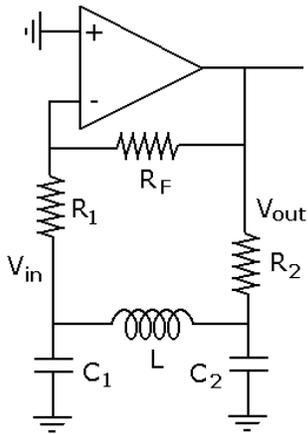}
\end{center}
\caption{\label{Fig4} A Colpitts oscillator that works right. } 
\end{figure}

There is though a simple way of solving this problem. This can be done by introducing a resistors ($R_2$) connected in series with the FDN as indicated in Fig.\ref{Fig4}. By doing so, $A$ remains unchanged but $F$ reads 

\begin{equation}
\label{Eq.3}
\noindent F(\omega)=\frac{R_1}
{[R_1(1-\omega^2LC_1)+j\omega L](1+j\omega R_2C_2)+R_2(1+j\omega R_1C_1)}\, .
\end{equation}

\noindent Thus, from $\Im{[F](\omega_0)}=0$ one obtains the oscillation frequency

\begin{equation}
\label{Eq.4}
\omega^2_0=\frac{C_1+C_2}{LC_1C_2}+\frac{1}{R_1R_2C_1C_2}\, ,
\end{equation}

\noindent and, since $A=-R_F/R_1$ one thus has

\begin{equation}
\label{Eq.5}
 \frac{R_FC_1C_2R_1R_2}{R_1R_2(R_1C_1^2+R_2C_2^2)+L(R_2C_1+R_1C_2)} \geq 1\, .
\end{equation}

\noindent In the case $R_1=R_2=R$ and $C_1=C_2=C$, Eqs. (\ref{Eq.4}-\ref{Eq.5}) reduce to 

\begin{equation}
\label{Eq.5b}
\omega^2_0=\frac{1}{LC}\left(2+\frac{L}{CR^2}\right)\, \mbox{\hspace{1cm} and \hspace{1cm}} 
R_F \geq 2R\left(1+\frac{L}{CR^2}\right) \, .
\end{equation}

It worth noticing that the results above differ from those obtained for a transconductance amplifier. The difference, however, is negligible small provided that $CR^2\gg L$. In fact, if $L/(CR^2)\ll$1, then, one has

\begin{equation}
\label{Eq.5c}
\omega^2_0 \approx \frac{2}{LC} \mbox{\hspace{1cm} and \hspace{1cm}} 
R_F \geq 2R \, .
\end{equation}

\noindent which are the results that normally appear in text-books for a Colpitts oscillator equipped with a transconductance amplifier.

In order to verify the results above, a sine-wave oscillator like that in Fig.\ref{Fig4} was mounted. A uA741 was used as voltage operational-amplifier and, $R$=1k$\Omega$, $C$=100 nF, $L$=3.9 mH and $R_F$=2.4 k$\Omega$. The oscillator was observed to oscillate at a frequency that was fairly close to the value predicted by Eq.(\ref{Eq.4}) and the amplitude of the oscillation was of the order of 1V (pp). Furthermore, the oscillation frequency was observed to be fairly insensitive to the power voltage from as low as $\pm$5 up to $\pm$15V.

\begin{figure}
\begin{center}
\epsfxsize=5cm\epsfbox{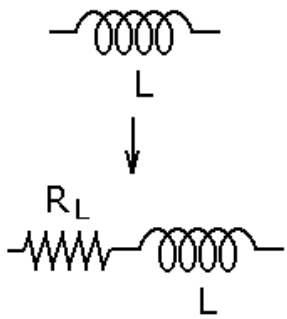}
\end{center}
\caption{\label{Fig5} Replacing an ideal by a more 'real' resistive inductor.} 
\end{figure}

For large $R$-values however, deviations from Eq.(\ref{Eq.5}) was observed. In fact, with increasing $R$ a $R_F$ larger than that predicted by Eq.(\ref{Eq.5}) appears to be required. This is found to be connected to the fact that for large $R$ the inductor resistance cannot be disregarded. This can be readily investigated by including the inductor resistance $R_L$ as indicated in Fig.\ref{Fig5} thus, after some algebra, one obtains

\begin{equation}
\label{Eq.6a}
\omega^2_0=\frac{1}{LC}\left[2\left(1+\frac{R_L}{R}\right)+\frac{L}{CR^2}\right]\, ,
\end{equation}

\noindent and,

\begin{equation}
\label{Eq.6b}
R_F \geq 2R\left[1+\frac{L}{CR^2}+\frac{2R_L}{R}+\frac{CRR_L}{L}\left(1+\frac{R_L}{R}\right)\right]\, . 
\end{equation}

\noindent It must be observed that, in the limiting case $R \gg R_L$, $L/(CR^2)\ll$1, and $CRR_L \ll L$, these equations reduce to those in (\ref{Eq.5c}). However, in the case that only the first two inequalities hold, then, in lieu of Eq.(\ref{Eq.6b}) one must use $R_F\geq 2R(1+CRR_L/L)$. For a $R_L$=8$\Omega$ inductor, as the one utilized in the present circuit, one has $R_F\geq 2R(1+R/5k\Omega)$. An equation that is observed to account for the present experiment fairly well.  

\begin{figure}
\begin{center}
\epsfxsize=8cm\epsfbox{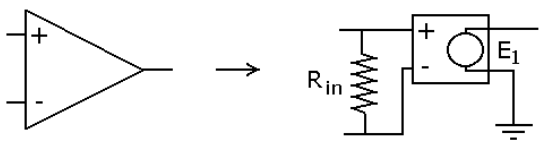}
\end{center}
\caption{\label{Fig6} A voltage-controlled source replaces the op-amp in the PSPICE simulations.} 
\end{figure}

PSPICE simulations of the circuit above have also been performed. To this end, the operational amplifier is replaced by a voltage-controlled source $E_1$ (see figure \ref{Fig6}) with an input resistance $R_{in}$ = 1M$\Omega$ and the same open-loop gain as that of the $\mu$A741 at the frequency of the oscillation. Similarly, an initial 0.01 V voltage is applied to capacitor $C_2$ since oscillations will not start up unless the circuit is somehow perturbed. The results of simulations show that the circuit delivers a neat, sinusoidal output voltage. A stable oscillation amplitude, however, is only attained after adjusting $R_F$ to 2.58 k$\Omega$, a value that compares fairly well with the minimum $R_F$ predicted by Eq.(\ref{Eq.6b}). 


In summary, it is found that LC-oscillators driven by a voltage operational-amplifier, as those that often appear in text-books of elementary electronics, may not necessarily work as expected. According to the present study, such a difficulty stems from the fact that the frequency-determining network (FDN) used in these circuits are designed for \emph{transconductance}, not for \emph{voltage} amplifiers. Curiously enough however, there is no warning about this point in such text-books and teachers may found it quite annoying at observing that, even though nothing seems to be wrong, the oscillator does not work as assumed. As is shown in this paper however, there seems to be a simple way of solving this problem. It consists in connecting a resistor in series with the FDN. By doing so, these LC-oscillators are observed to work as is expected from the block diagram in Fig.\ref{Fig1} and the Barkhausen criteria in Eqs.(\ref{Eq.1}). It must be also mentioned that the so-called Hartley oscillator exhibited the same kind of problem. Again, such a difficulty can be readily solved by adding a resistor in series with the FDN as discussed above.

\end{document}